# Comments on "Ballistics: a primer for the surgeon"

Michael Courtney
Ballistics Testing Group, P.O. Box 24, West Point, NY 10996
Michael_Courtney@alum.mit.edu

Amy Courtney
Department of Physics, United States Military Academy, West Point, NY 10996

**Abstract:**
In response to a published assertion to the contrary, this paper briefly reviews many studies that document remote wounding effects of ballistic pressure waves including experiments in pigs and dogs that find brain injury resulting from animal models shot in the thigh and case studies in humans that document both remote brain and spinal cord injuries ascribed to ballistic pressure waves.

The authors of "Ballistics: a primer for the surgeon"[17] write:

*In addition to the transverse, low-frequency wave that emanates from the projectile as it passes through tissue, which creates the temporary cavity, there are also high-frequency, low-amplitude stress waves that are produced in front of the projectile. These compressive waves are similar to the shock waves seen in front of high-speed aircraft. It is not clear what effect these waves have on wounding characteristics of a projectile, but it is likely that they play a small role in producing permanent injury.*

In contrast to this assertion, many studies document remote wounding effects of ballistic pressure waves. Early studies revealed that the magnitudes of the oscillating high-frequency ballistic pressure wave could be over 1000 kPa (147 psi)[3] and documented wounding effects in peripheral nerves.[4,7]

A series of experiments[15,10,11,12,13] studied pigs shot in the thigh. High-speed pressure transducers were implanted in the thigh, abdomen, neck, and brain. These experiments determined that ballistic pressure waves over 400 kPa (59 psi) reach the contralateral thigh and waves as large as 300 kPa (44 psi) can reach the brain.[12] Electron microscopy revealed damage in peripheral nerves, spinal cord, and brain. A similar experiment[18] in dogs found significant damage in both the hypothalamus and hippocampus regions of the brain due to remote effects of the ballistic pressure wave. A recent paper reviews links between traumatic brain injury and ballistic pressure waves.[1]

Case studies in human patients also document pressure wave effects in the central nervous system. In 1941, a soldier received a bullet wound near but not directly impacting the brain.[16] The patient experienced acute epileptic symptoms which reappeared nearly 50 years later with secondary generalisation. The authors attribute this pathology to the pressure wave.

Several studies reveal remote pressure wave wounding in the spinal cord.[8,9,14] For example, a woman shot with a 9mm handgun experienced specific neurological symptoms in the spine remote from the wound channel.[9] This wounding was explained by focusing of the pressure wave on the spinal cord (similar to the way a lens can focus light waves).

Finally, the formula given for drag force[17] is incorrect and not found in the citation.[6] The drag force magnitude is:[2,5]

$$F_d = \frac{1}{2} C_d \times d \times A \times v^2,$$

where $F_d$ is the drag force, $C_d$ is the drag coefficient, d is the fluid density, A is the cross sectional area of the projectile, and v is the projectile velocity.

## About the Authors

*Amy Courtney* currently serves on the Physics faculty of the United States Military Academy at West Point. She earned a MS in Biomedical Engineering from Harvard University and a PhD in Medical Engineering and Medical Physics from a joint Harvard/MIT program. She has published work in orthopedic biomechanics, biomedical engineering, ballistics, acoustics, medical physics, and traumatic brain injury. She has taught Anatomy and Physiology as well as Physics. She has served as a research scientist at the Cleveland Clinic and Western Carolina University, as well as on the Biomedical Engineering faculty of The Ohio State University.

*Michael Courtney* earned a PhD in experimental Physics from the Massachusetts Institute of Technology. He has published work in theoretical astrophysics, theoretical and experimental atomic physics, chaos, ballistics, acoustics, medical physics, and traumatic brain injury. He has also served as the Director of the Forensic Science Program at Western Carolina University and also been a Physics Professor, teaching Physics, Statistics, and Forensic Science. Michael and his wife, Amy, founded the Ballistics Testing Group in 2001 to study incapacitation ballistics and the reconstruction of shooting events.